\documentclass[final,5p,times,twocolumn,authoryear]{elsarticle}

\usepackage{amsmath, amssymb}
\usepackage{lmodern}
\usepackage{latexsym}
\usepackage{amsmath}
\usepackage{amssymb}
\usepackage{amsbsy}
\usepackage{amsthm}
\usepackage{amsfonts}
\usepackage{mathrsfs}
\usepackage[cal=cm,scr=boondoxo]{mathalfa}
\usepackage{bm}
\usepackage{graphicx}
\usepackage{sansmath}
\usepackage{relsize}
\usepackage{graphicx}
\usepackage[utf8]{inputenc} 
\usepackage{xcolor}
\usepackage[T1]{fontenc}
\usepackage{epstopdf}
\usepackage{esdiff}
\usepackage[T1]{fontenc}
\usepackage{ae,aecompl}
\usepackage{mathtools}
\usepackage{array}
\usepackage{booktabs}
\usepackage{longtable}
\usepackage{appendix}
\usepackage{hyperref}
\usepackage[mathlines]{lineno}

\journal{Journal of High Energy Astrophysics}

\usepackage{amssymb}
\usepackage{lipsum}

\journal{Journal of High Energy Astrophysics}

\begin{document}

\begin{frontmatter}

\title{The Role of r-Modes in Pulsar Spin-down, Pulsar Timing, and Gravitational Waves}

\author[UWO1]{Xiyuan Li}

\author[UWO1]{Shahram Abbassi\corref{cor1}}
\ead{sabbassi@uwo.ca}
\cortext[cor1]{Corresponding author}

\author[IITH]{Varenya Upadhyaya}

\author[UWO2]{Xiyang Zhang}

\author[UWO1,UWO3]{S.~R. Valluri}

\address[UWO1]{Department of Physics and Astronomy, University of Western Ontario, London, ON N6A 3K7, Canada}
\address[UWO2]{Department of Statistical and Actuarial Sciences, University of Western Ontario, London, ON N6A 3K7, Canada}
\address[UWO3]{Department of Mathematics, King’s University College, University of Western Ontario, London, ON N6A 3K7, Canada}
\address[IITH]{Department of Physics, Indian Institute of Technology, Hyderabad, Telangana 502284, India}

\begin{abstract}
We investigate the role of r-mode oscillations in pulsar spin-down and their implications for gravitational wave emission and pulsar timing analysis. Using a non-linear differential framework that includes r-mode contributions, we derive time-dependent solutions for rotational frequency and period evolution. These expressions are validated using observational data from the Crab pulsar with high precision. By analytically fitting braking indices and spin-down coefficients, we link measurable pulsar properties to gravitational wave signatures. Furthermore, we present closed-form expressions for neutron star compactness and tidal deformability using Lambert W and Lambert–Tsallis functions, enabling model-independent inferences from r-mode gravitational wave frequencies. Our results show that incorporating r-modes significantly improves the accuracy of spin-down models and continuous wave detectability, particularly through the inclusion of high-order frequency terms. This framework supports the modeling of timing residuals, glitch quantification, and gravitational wave constraints. Our findings have direct relevance for data analysis in ongoing and future gravitational wave observatories.
\end{abstract}

\begin{keyword}
pulsars \sep neutron stars \sep gravitational waves \sep r-modes \sep spin-down \sep timing \sep tidal deformability
\end{keyword}

\end{frontmatter}



\begin{keyword}
pulsars: general --- gravitational waves --- neutron stars --- r-modes --- spin-down --- tidal deformability --- compactness --- continuous wave detection --- astrophysical fluid dynamics



\end{keyword}





\section{Introduction} 
\label{sec:intro}

Neutron stars are among the most extreme astrophysical laboratories, hosting ultradense matter, strong magnetic fields, and rapid rotation. Pulsars—magnetized rotating neutron stars emitting periodic electromagnetic signals—lose angular momentum through various channels, including electromagnetic radiation, particle winds, and gravitational wave (GW) emission. The timing of pulsars, especially millisecond pulsars acting as highly stable cosmic clocks, provides powerful probes of these processes \citep{LorimerHandbook2004, ProgressUpdate_Taylor, TaylorClocks1991, MatsakisClocks1997}.

A theoretically compelling mechanism for GW emission from rotating neutron stars involves the excitation of r-modes—toroidal, inertial oscillations restored by the Coriolis force and analogous to Rossby waves in terrestrial atmospheres \citep{1978MNRAS.184..501P, 1981A&A....94..126P}. These modes become unstable via the Chandrasekhar–Friedman–Schutz (CFS) mechanism when gravitational radiation drives the mode amplitude to grow faster than internal viscous dissipation can suppress it \citep{1998MNRAS.299.1059A, 1998ApJ...502..714F, Andersson_1998, Abbassi2012}. Once excited, r-modes can dominate angular momentum loss and significantly alter the spin evolution of young or rapidly rotating neutron stars.

The dissipation of r-modes is predicted to emit continuous, quasi-monochromatic gravitational waves, a key target for detectors such as LIGO, Virgo, and KAGRA \citep{Abbott2016_GW150914, KAGRA.2019}. While the first detections (e.g., GW150914) involved compact binary mergers \citep{Abbott2016_GW150914, PhysRevLett.120.031104}, recent targeted searches have placed upper limits on r-mode amplitudes in individual pulsars such as PSR J0537–6910 \citep{LIGO_rmodes_Abbott_2021}. The r-mode instability has also been proposed as an explanation for the observed spin-down in the Crab pulsar \citep{1998PhRvD..58h4020O, Strohmayer2013}, and its evolution is known to be sensitive to factors such as crust elasticity, magnetic fields, differential rotation, and superfluidity \citep{2000ApJ...529L..33B, 2002MNRAS.337.1224A, 2012MNRAS.424...93H, 2002ApJ...574..899R}.

With next-generation detectors—Einstein Telescope \citep{branchesi2023science}, Cosmic Explorer \citep{evans2021horizon}, LISA \citep{amaroseoane2017laser}, Taiji \citep{Ruan2020}, and Tianqin \citep{Luo2016} —the detection prospects for continuous waves from r-mode activity will improve significantly. Although low-frequency space-based detectors operate primarily in the nanohertz to millihertz range, they will inform high-frequency models by refining estimates of orbital dynamics and neutron star mass distributions. In parallel, Pulsar Timing Arrays (PTAs)—including NANOGrav, EPTA, PPTA, and the future Square Kilometer Array—offer complementary sensitivity to ultra-low-frequency GW signals by analyzing timing residuals across ensembles of millisecond pulsars \citep{Nanograv, Manchester_2013_ipta, Hobbs_2010, Weltman_2020}.

Despite these advances, several key questions remain unresolved. What is the precise impact of r-modes on observable spin-down behavior? Can gravitational wave signals from r-modes be detected within realistic sensitivities? And can such signals be used to constrain fundamental neutron star properties like the equation of state (EoS), compactness, and tidal deformability?

This work addresses these questions by developing a unified, analytical model that incorporates r-mode contributions to pulsar spin-down alongside electromagnetic and quadrupole GW torques. Using Lambert W and Lambert–Tsallis functions, we derive closed-form relationships connecting r-mode gravitational wave frequency to neutron star compactness and tidal deformability, generalizing previous approaches \citep{Strohmayer2013, Ghosh_2023}. Our framework is EoS-independent, time-resolved, and applicable across the pre- and post-glitch evolution phases.

We validate our model using Crab pulsar data, demonstrating high-accuracy agreement in period evolution and braking indices. In particular, we show that including r-mode contributions improves spin-down modeling, enhances sensitivity to timing irregularities, and provides a pathway to constrain gravitational wave emission from isolated neutron stars.

The paper is structured as follows: Section~\ref{sec:spindown} introduces the r-mode-augmented spin-down torque model. Section~\ref{sec:analysis} develops analytical relations for rotational frequency, r-mode amplitude, compactness, tidal deformability, and pulsar period evolution. Section~\ref{sec:period modeling} focuses on period modeling and braking indices. Section~\ref{sec:results} compares the model to observational data, with emphasis on glitch behavior and timing residuals. Section~\ref{sec:disc} discusses astrophysical implications, and Section~\ref{sec:sum} concludes. Detailed derivations are provided in Appendices~\ref{Appendix: freq analysis} to~\ref{sec:appendix}.

\section{Spin-down Mechanisms}
\label{sec:spindown}

Pulsars lose rotational energy over time due to several dissipative mechanisms, with magnetic dipole radiation typically dominating the long-term spin evolution of isolated neutron stars \cite{1968Natur.218..731G, 1968Natur.219..145P}. However, for younger or faster-spinning pulsars, gravitational wave emission—especially through unstable oscillatory modes such as r-modes—can contribute significantly to the spin-down torque. This section analyzes how r-mode energy loss alters pulsar spin-down and connects with gravitational wave observables \cite{1969ApJ...157.1395O, 1969ApJ...158L..71F, papaloizou, Andersson_1998}.

Recent work has introduced simplified expressions for damping timescales associated with shear and bulk viscosity, both of which play critical roles in regulating r-mode growth and saturation \cite{2003ApJ...591.1129A, Lindblom2003}.The cumulative energy loss is inferred by considering the total change in rotational energy from all dominant channels \cite{refId0}:
\begin{align}
    \dot{E}_{\text{mono}} &= -\beta \frac{\mu}{cR^2}\Omega^2, \\
    \dot{E}_{\text{dip}}  &= -\frac{2\mu^2}{3c^3}\Omega^4, \\
    \dot{E}_{\text{quad}} &= -\frac{32GI^2e^2}{5c^5}\Omega^6, \\
    \dot{E}_{\text{r-mode}} &= -\frac{32 \pi^2 \alpha^2 J M R^3 \Omega^8}{c^7},
\end{align}
where $\Omega = 2\pi \nu$ is the angular frequency, $\mu$ is the magnetic dipole moment, $R$ is the stellar radius, $I$ the moment of inertia, $e$ the ellipticity, and $\alpha$ the r-mode amplitude. $J$ is a structural constant related to the stellar current quadrupole moment. The form of $\dot{E}_{\text{r-mode}}$ aligns with canonical derivations for unstable current-quadrupole r-mode dissipation \cite{Lindblom_1998, 1998PhRvD..58h4020O, Andersson_1999}.

For millisecond pulsars ($P < 1.5$ ms), variations in the moment of inertia may arise from internal structural changes, including phase transitions \cite{PhysRevLett.79.1603, Alford2014}. However, for slower rotators ($P \gtrsim 3$ ms), such variations are negligible and $I$ can be treated as constant \cite{refId0}.

The spin frequency evolution can be connected to energy loss using:
\begin{equation}
\begin{aligned}
    E_{\text{rot}} &= \frac{1}{2}I\Omega^2, \\
    \dot{\nu} &= \frac{1}{4\pi^2 I} \cdot \frac{\dot{E}_{\text{rot}}}{\nu}.
    \label{eq:f-energy}
\end{aligned}
\end{equation}

Equation~\eqref{eq:f-energy} forms the basis for modeling multi-channel torque effects. A generic spin-down model takes the form:
\begin{equation}
    \dot{\nu} = k \nu^n,
    \label{eq:spindown_braking}
\end{equation}
where $k$ is a constant and $n$ is the braking index, which reflects the dominant energy loss process. To capture additional physical processes, including gravitational wave emission and r-mode losses, we generalize the model as:
\begin{equation}
    \dot{\nu} = -s(t)\nu - r(t)\nu^3 - g(t)\nu^5 - l(t)\nu^7,
    \label{eq:spindown_coeff}
\end{equation}
where $s$, $r$, $g$, and $l$ represent frequency-dependent coefficients corresponding to monopole, dipole, quadrupole, and r-mode current-quadrupole losses, respectively.

Each term in Eq.~\eqref{eq:spindown_coeff} corresponds to a specific physical mechanism: magnetic monopole ($\nu^1$), dipole radiation ($\nu^3$), quadrupole deformation ($\nu^5$), and r-mode emission ($\nu^7$), which maps to an $n = 8$ braking index. Such a steep frequency dependence is characteristic of current-quadrupole gravitational radiation. The effect of these terms becomes pronounced in young, hot neutron stars or those with substantial internal fluid motion \cite{1998PhRvD..58h4020O, Andersson_1999}.

In addition to axisymmetric deformations, high-order multipoles such as octupoles or baroclinic modes may be present in certain exotic systems, although these are not included in our model \cite{Pani.octupole.PhysRevD.92.124003, Mastrano.octupole.8184360}.

Ultimately, r-mode-driven gravitational radiation growth is governed by the CFS mechanism \cite{Chandrasekhar1970, FriedmanSchutz}. When r-mode growth timescales are shorter than viscous damping, the mode becomes unstable, causing a strong angular momentum outflow via gravitational waves. This process introduces a measurable spin-down signal. Numerous works have modeled the associated frequencies and detection strategies \cite{Lindblom_1998, Lindblom_2001, Andersson_1999, Ghosh_2023, Idrisy_2015, Abbott_2021, Abbott2021b}.

In the next section, we incorporate these mechanisms into analytical and numerical models to study compactness, deformability, r-mode amplitudes, and spin-down behavior.

\section{r-mode Formalism and Analytical Solutions}
\label{sec:analysis}
\subsection{Rotational and Gravitational Wave Frequencies from r-modes}
The r-mode oscillation frequency $\omega$ in a rotating neutron star is governed by its angular frequency $\Omega$ and the mode's angular indices $l$ and $m$, given by:
\begin{equation}
    \omega = -m\Omega + \frac{2m\Omega}{l(l+1)},
\end{equation}
where $l$ and $m$ specify the spherical harmonic degree and azimuthal order, respectively \cite{1981A&A....94..126P, Andersson_2003a}. We consider the fundamental $l = m = 2$ r-mode, which dominates gravitational radiation among inertial modes \cite{Lockitch_1999, Yoshida_2000, Yoshida.2005.mnras}. This analysis assumes barotropic stellar structure, neglecting buoyancy effects associated with the Brunt–Väisälä frequency. While this enables analytical tractability, future studies may incorporate stratification for a more realistic model of neutron star interiors.

For $l = m$, the r-mode angular frequency simplifies to:
\begin{equation}
    \omega = -\frac{(l+2)(l-1)}{l+1}\Omega,
\end{equation}
and the corresponding gravitational wave frequency is:
\begin{equation}
    f_{gw} = -\frac{2}{3\pi}\Omega.
    \label{eq:linear r-mode freq}
\end{equation}

Equation~\eqref{eq:linear r-mode freq} is valid in the Newtonian regime. However, neutron stars are compact and rapidly rotating, necessitating general relativistic (GR) corrections. For this, we adopt a higher-order approximation \cite{Caride.PhysRevD.100.064013, Ghosh_2023}:
\begin{equation}
     f_{gw} = A\nu - \frac{B}{\nu_k^2}\nu^3,
    \label{eq:trinomial}
\end{equation}
where $\nu$ is the spin frequency (Hz), $\nu_k$ is the Keplerian breakup frequency, the upper spin limit for stable pulsar rotation that can be determined analytically,\citep{Levin.2001MNRAS.324..917L, 1998PhRvD..58h4020O}. A typical $\nu_k$ for a neutron star on the lower-end of the mass range is $~506$ Hz \cite{Caride.PhysRevD.100.064013}. The parameters $A$ and $B$ encapsulate GR and rotational corrections, respectively:
\begin{equation}
\begin{aligned}
    1.39 \leq &A \leq 1.57,\\
    0 \leq &B \leq 0.195.
    \label{val: A and B}
\end{aligned}
\end{equation}

Values of $A$ reflect frame dragging and spacetime curvature in slowly rotating stars, while $B$ accounts for centrifugal flattening and nonlinear rotational effects \cite{Yoshida.2005.mnras, Caride.PhysRevD.100.064013, Abbott_2021}.

Equation~\eqref{eq:trinomial} is non-linear in $\nu$, making analytical inversion difficult. We resolve this by applying the Lambert–Tsallis function, a generalization of the Lambert W function suitable for non-linear systems. The $q$-exponential used in the Tsallis formalism is defined as:
\begin{equation}
    \exp_q{z} = \left[1+(1-q)z\right]^{\frac{1}{1-q}}, \quad q \neq 1.
\end{equation}
The Lambert–Tsallis function $W_q(z)$ satisfies:
\begin{equation}
    W_q(z)\exp_q{W_q(z)} = z, \quad z \in \mathbb{R}.
\end{equation}

Solving Eq.~\eqref{eq:trinomial} with this approach yields:
\begin{align}
    \nu_1 &= \left[-\frac{A}{2B}\nu_k^2 W_{1/2}\left\{-\frac{2B}{A\nu_k^2}\left(\frac{f_{gw}}{A}\right)^2\right\}\right]^{1/2},
    \label{eq: Lambert-Tsallis nu_1} \\
    \nu_2 &= \left[\frac{3B}{2A\nu_k^2} W_{5/2}\left\{\frac{2A\nu_k^2}{3B}\left(\frac{-f_{gw}\nu_k^2}{B}\right)^{-2/3}\right\}\right]^{-1/2}.
    \label{eq: Lambert-Tsallis nu_2}
\end{align}

These solutions offer a path to estimate $\nu$ from $f_{gw}$ analytically, bypassing numerical inversion. For PSR J0537–6910, a young pulsar with $\nu = 62$ Hz \cite{1998ApJ...499L.179M}, inserting $A$ and $B$ into Eq.~\eqref{eq:trinomial} gives an r-mode GW frequency of 86–98 Hz—well within LIGO's sensitivity band \cite{LIGO_rmodes_Abbott_2021, Ghosh_2023}. From Eq.~\eqref{eq: Lambert-Tsallis nu_1}, we find the rotational frequency range $62.00 \leq \nu_1 \leq 62.42$ Hz, consistent with expected values. However, $\nu_2$ from Eq.~\eqref{eq: Lambert-Tsallis nu_2} lies outside physical bounds and is excluded.

This analytic method provides an elegant way to probe the spin evolution of neutron stars via gravitational wave frequency measurements and can be extended to study braking indices and timing residuals.

\subsection{Structural Inference: Compactness and Tidal Deformability from r-modes}

The compactness of a neutron star, defined as $C = M/R$, influences its r-mode gravitational wave frequency. A phenomenological model that captures this relationship is given by:
\begin{equation}
    f_r = a_1 \left(\frac{M}{R}\right) - a_2 \left(\frac{M}{R}\right)^2 + a_3,
    \label{eq:compactness_rmode_freq}
\end{equation}
where $a_1$, $a_2$, and $a_3$ are empirical coefficients fitted using tabulated EoS models \cite{Idrisy_2015, Ghosh_2023}. This quadratic form provides an intuitive interpolation between r-mode frequency and compactness.

By approximating Eq.~\eqref{eq:compactness_rmode_freq} in the limit of small $C$, we obtain:
\begin{equation}
    f_r - a_3 = -a_2 C^2 \exp\left\{-\frac{a_1}{a_2} \cdot \frac{1}{C}\right\},
    \label{eq:compactness_approx}
\end{equation}
which can be analytically inverted using the Lambert W function to yield:
\begin{equation}
    C = \frac{a_1}{2a_2} \left(W\left\{ \frac{a_1}{2a_2} \sqrt{\frac{a_2}{a_3 - f_r}} \right\}\right)^{-1},
    \label{eq:compactness Lambert}
\end{equation}
where $W(z)$ is the Lambert W function satisfying $W(z)\exp\{W(z)\} = z$ \cite{Corless.1996}.

Although Eq.~\eqref{eq:compactness_rmode_freq} appears independent of spin frequency, the compactness and tidal deformability implicitly contain rotational information via their dependence on stellar structure. As shown in \cite{Ghosh_2023}, the r-mode frequency encapsulates rotational and structural dynamics, enabling these quantities to be inferred from GW observations.

\begin{figure}
    \centering
    \includegraphics[width=0.45\textwidth]{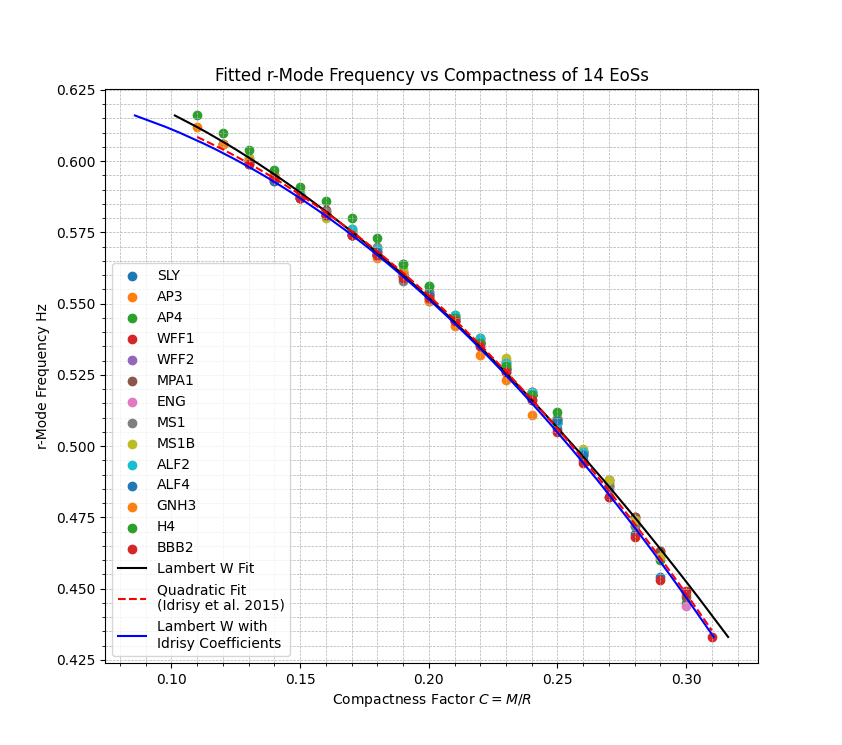}
    \caption{Fitted r-mode gravitational wave frequency versus compactness \( C = M/R \) for 14 tabulated equations of state (EoSs) \cite{Idrisy_2015}. Solid black line: Lambert W solution given by Eq.~\eqref{eq:compactness Lambert}, with best-fit coefficients \( a_1 = -0.184 \), \( a_2 = 1.640 \), and \( a_3 = 0.667 \), achieving an \( R^2 \) value of 0.9972. Dashed red line: Quadratic model, with best-fit coefficients \( a_1 = 0.079 \), \( a_2 = 2.25 \), and \( a_3 = 0.627 \) from Idrisy et al. 2015 \cite{Idrisy_2015}. Solid blue line: Lambert W solution with Idrisy quadratic model coefficients.} The Lambert W fit provides a superior match to the numerical data, capturing the nonlinear dependence of r-mode frequency on compactness across a wide range of realistic neutron star models.

    \label{fig:compactness}
\end{figure}

In Figure~\ref{fig:compactness}, we fit Eq.~\eqref{eq:compactness Lambert} to data from 14 tabulated EoSs \cite{Idrisy_2015}. The best-fit parameters—$a_1 = -0.184$, $a_2 = 1.640$, $a_3 = 0.667$—yield $R^2 = 0.9972$, confirming the robustness of the Lambert W solution. We also evaluated the frequency estimation performance of the quadratic model and the Lambert W model using the same set of best-fit coefficients from Idrisy et al. 2015 \cite{Idrisy_2015}. The maximum difference between the two estimated frequencies is within $0.34\%$. Equation \ref{eq:compactness Lambert} provides a closed-form expression for estimating neutron star compactness from r-mode GW frequency, independent of the EoS.

Tidal deformability $\Lambda$ characterizes a star's response to tidal perturbations and also affects r-mode evolution. It is defined as:
\begin{equation}
    \Lambda = \frac{2}{3}k_2 \left(\frac{R}{M}\right)^5,
\end{equation}
where $k_2$ is the dimensionless Love number \cite{Flanagan_2008, Hinderer_2008}. The r-mode frequency is indirectly influenced by $\Lambda$ through changes in the star’s density distribution and moment of inertia.

A quadratic model relating $f_r$ to $\ln \Lambda$ is:
\begin{equation}
    f_r = b_1 \ln \Lambda + b_2 (\ln \Lambda)^2 + b_3,
    \label{eq:tidal deformability}
\end{equation}
where $b_1$, $b_2$, $b_3$ are fitted coefficients \cite{gupta2022determining}.Under the assumption that $b_2 < 0$ \cite{gupta2022determining, Ghosh_2023}, and $\frac{-b_2}{b_1} \ln \Lambda \ll 1$, Eq.~\eqref{eq:tidal deformability} becomes:
\begin{equation}
    \frac{b_2}{b_1^2}(f_r - b_3) = \frac{b_2}{b_1} \ln \Lambda \cdot \exp\left\{\frac{b_2}{b_1} \ln \Lambda\right\},
    \label{eq:lambda approximation}
\end{equation}
which has an analytical solution via the Lambert W function:
\begin{equation}
    \ln \Lambda = \frac{b_1}{b_2} W\left\{\frac{b_2}{b_1^2}(f_r - b_3)\right\}.
    \label{eq:tidal deformability Lambert}
\end{equation}

\begin{figure}
    \centering
    \includegraphics[width=0.45\textwidth]{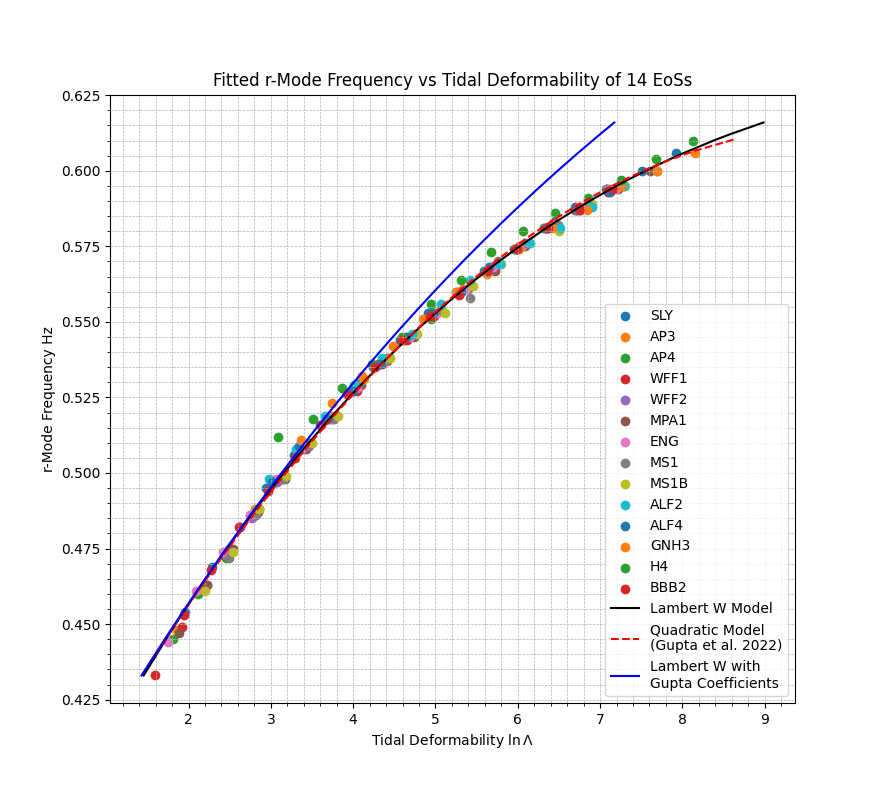}
    \caption{Fitted r-mode gravitational wave frequency as a function of tidal deformability \( \ln \Lambda \) for 14 tabulated EoSs \cite{Idrisy_2015}, computed using LALSuite \cite{2020ascl.soft12021L}. Solid black line: Lambert W fit (Eq.~\eqref{eq:tidal deformability Lambert}) with best-fit coefficients \( b_1 = 0.0568 \), \( b_2 = -0.0043 \), and \( b_3 = 0.3591 \), achieving an \( R^2 \) value of 0.998. Dashed red line: Quadratic model, with best-fit coefficients \( b_1 = 0.0498 \), \( b_2 = -0.0025\), and \( b_3 = 0.3668 \) from Gupta et al. 2022 \cite{gupta2022determining}. Solid blue line: Lambert W solution with Gupta quadratic model coefficients.} The Lambert W model provides a highly accurate, analytical, EoS-independent mapping between tidal deformability and r-mode frequency.

    \label{fig:deformability}
\end{figure}

Figure~\ref{fig:deformability} shows the fit of Eq.~\eqref{eq:tidal deformability Lambert} to data obtained using the TOV solver in LALSuite \cite{2020ascl.soft12021L}. Compactness inputs are taken from \cite{Idrisy_2015}. Although $\nu$ does not appear explicitly, its influence is implicit through compactness and structural changes. The best-fit values—$b_1 = 0.0568$, $b_2 = 0.0043$, $b_3 = 0.3591$—achieve $R^2 = 0.998$, demonstrating excellent agreement. When our Lambert W model and the Gupta-like quadratic model were evaluated using the same set of best-fit coefficients from Gupta et al. 2022 \cite{gupta2022determining}, the maximum difference between the two estimated frequencies is within $3.25 \%$.

\subsection{Time-Dependent Amplitude and Energy Loss}
\label{subsec:lambert}

The evolution of r-mode amplitude $\alpha$ and stellar spin $\Omega$ in a rotating neutron star can be modeled as a coupled two-variable system. Following the formalism of \cite{PhysRevD.58.084020}, we begin with:
\begin{align}
    \frac{d\alpha}{dt} &= -\frac{\alpha}{\tau_{GR}} - \frac{\alpha}{\tau_{V}} \cdot \frac{1 - \alpha^2 Q}{1 + \alpha^2 Q},
    \label{eq: dalphadt} \\
    \frac{d\Omega}{dt} &= -\frac{2\Omega}{\tau_V} \cdot \frac{\alpha^2 Q}{1 + \alpha^2 Q},
    \label{eq:r-mode amplitude freq}
\end{align}
Here, $Q = 3\tilde{J}/2\tilde{I}$ is a dimensionless structural parameter dependent on the equation of state. $\tilde{J}$ and $\tilde{I}$ are normalized stellar structure coefficients, while $\tau_{GR}$ and $\tau_V$ are the gravitational radiation and viscous damping timescales, respectively \cite{Lindblom_1998}.

Equation~\eqref{eq: dalphadt} can be rewritten in a more tractable form:
\begin{equation}
    \frac{\tau_{GR} \tau_V (1 + \alpha^2 Q) \, d\alpha}{\alpha\left[(\tau_V + \tau_{GR}) + \alpha^2 (\tau_V Q - \tau_{GR} Q)\right]} = -dt,
    \label{eq:rearranged dalphadt}
\end{equation}
Defining:
\begin{equation*}
    \xi_1 = \tau_{GR} \tau_V, \quad T_A = \tau_{GR} + \tau_V, \quad T_B = Q(\tau_V - \tau_{GR}),
\end{equation*}
we simplify Eq.~\eqref{eq:rearranged dalphadt} as:
\begin{equation}
    \frac{\xi_1 (1 + \alpha^2 Q) \, d\alpha}{\alpha (T_A + \alpha^2 T_B)} = -dt.
    \label{eq:timescale substituted dalphadt}
\end{equation}

Upon integrating and inverting the expression using the Lambert $W$ function, the amplitude $\alpha$ becomes:
\begin{equation}
    \alpha(t) = \left(\frac{1}{2\xi_c} W\left[2\xi_c \exp\left(-\frac{2(t + \xi_B \ln T_A - c_1)}{\xi_A}\right)\right]\right)^{1/2},
\end{equation}
where the intermediate coefficients are:
\begin{align*}
    \xi_A &= \xi_1 / T_A, \\
    \xi_B &= \frac{Q \xi_1}{2 T_B} - \frac{\xi_1}{2 T_A}, \\
    \xi_c &= \frac{T_B \xi_B}{T_A \xi_A}.
\end{align*}

The corresponding spin frequency is given by $\nu = \Omega / (2\pi)$, which varies with time according to Eq.~\eqref{eq:r-mode amplitude freq}.
\begin{figure}
    \centering
    \includegraphics[width=0.45\textwidth]{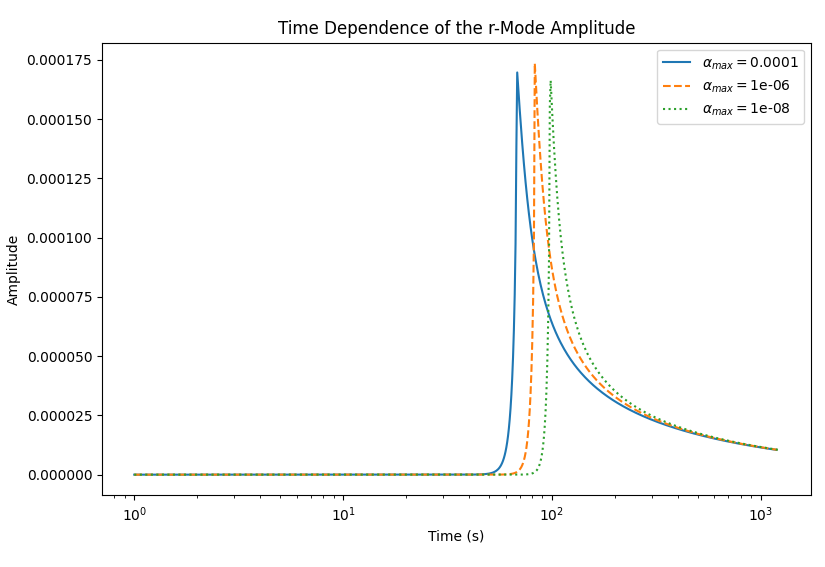}
    \caption{Time dependence of the r-mode amplitude $\alpha$. The amplitude grows into the saturation mode near $t=10^2$ s, where the non-linear effects can no longer be ignored. The r-mode amplitude settles to a gradual decay that dominates the rest of the time.}
    \label{fig:alpha 28}
\end{figure}

To connect with observations, we relate the r-mode amplitude $\alpha$ to the gravitational wave strain $h_0$ via \cite{Abbott_2021}:
\begin{equation}
    \alpha = \left(\frac{5}{8\pi}\right)^{1/2} \frac{c^5}{G} \cdot \frac{h_0}{(2\pi f_{gw})^3} \cdot \frac{d}{M R^3 \tilde{J}},
    \label{eq:alpha_h0}
\end{equation}
where $f_{gw}$ is the gravitational wave frequency, $d$ is the source distance, $M$ and $R$ are the mass and radius of the neutron star, and $\tilde{J}$ reflects the current quadrupole normalization.

These expressions link the r-mode amplitude and its evolution to gravitational wave observables. The inclusion of the Lambert $W$ function enables analytical tracking of $\alpha(t)$ across viscous and radiative timescales, offering a predictive framework for gravitational wave data analysis in the context of r-mode-driven signals. 

Furthermore, our amplitude formulation is consistent with observational upper limits reported by \citep{Strohmayer2013}, who found that r-mode amplitudes in sources such as the Crab pulsar are likely bounded by $\alpha \lesssim 10^{-4}$. These empirical constraints provide essential context for interpreting Eq.~\eqref{eq:alpha_h0}, particularly in relation to gravitational wave detectability using current and future detectors.

\section{Pulsar Period Modeling and Spin-down Coefficients}
\label{sec:period modeling}
\subsection{Period Evolution with Time-Dependent Spin-down Parameters}
\label{subsec:period}

We extend our spin-down model by deriving a closed-form expression for the time evolution of the pulsar period $P(t)$, following the non-linear formalism of \cite{Chishtie_2018}. The model applies to isolated pulsars and incorporates time-dependent spin-down coefficients $\{s(t), r(t), g(t), l(t)\}$, as introduced in Eq.~\eqref{eq:spindown_coeff}.

The model's consistency is validated against observed frequency derivatives and braking indices from \cite{10.1093/mnras/233.3.667}, and with data from the ATNF Pulsar Database. For the Crab pulsar, our predicted periods match observed values with a relative error of 0.02\%, as shown in Table~\ref{tab:1}.

Rewriting Eq.~\eqref{eq:spindown_coeff} in terms of the period $P = 1/f$, where $f$ is the pulsar frequency, and factoring out $P^5$, yields:
\begin{equation}
    P^5 \frac{dP}{dt} = \left(s_0 P^6 + r_0 P^4 + g_0 P^2 + l_0\right) \left(1 + \frac{t}{t_c}\right)^{-2},
    \label{eq:P^5}
\end{equation}
where $t_c$ is the magnetic field decay timescale. Defining $K = P^2$ and using $dK = 2P dP$, Eq.~\eqref{eq:P^5} becomes:
\begin{equation}
    \frac{K^2 dK}{2s_0 [(K - a)(K - b)(K - c)]} = \left(1 + \frac{t}{t_c}\right)^{-2} dt,
    \label{eq:period and time}
\end{equation}
where $a$, $b$, $c$ are the roots of the spin-down polynomial.
Integrating Eq.~\eqref{eq:period and time} yields:
\begin{multline}
    2[(x-a)(x^2 - y^2) + 2xy^2] \tan^{-1}\left(\frac{y(K_0 - K)}{(K - x)(K_0 - x) + y^2}\right) + \\
    [y(x^2 - y^2) - 2xy(x - a)] \ln\left(\frac{(K - x)^2 + y^2}{(K_0 - x)^2 + y^2}\right) \\
    - 2ya^2 \ln\left(\frac{K - a}{K_0 - a}\right) = -4ys_0 t(x^2 - 2ax + a^2 + y^2),
    \label{eq:period_implicit}
\end{multline}
where $x$ and $y$ are defined through the complex root structure of the cubic. Table~\ref{tab:crab_approx1} provides term-by-term estimates for the Crab pulsar.

\begin{table}[htbp]
\centering
\renewcommand{\arraystretch}{1.2}
\setlength{\tabcolsep}{6pt}
\begin{tabular}{@{}l r@{}}
\hline\hline
Function & Value (Crab PSR B0531+21) \\
\hline
$\ln\left[\frac{(K - x)^2 + y^2}{(K_0 - x)^2 + y^2}\right]$ & 0.0158 \\
$\tan^{-1}\left[\frac{y(K_0 - K)}{(K - x)(K_0 - x) + y^2}\right]$ & $-0.066$ \\
$\ln\left[\frac{K - a}{K_0 - a}\right]$ & 0.0540 \\
\hline\hline
\end{tabular}
\caption{\label{tab:crab_approx1}Approximate values for logarithmic and arctangent terms in Eq.~\eqref{eq:period_implicit} for the Crab pulsar, enabling Taylor expansion.}
\end{table}

Using Taylor expansion for small deviations, Eq.~\eqref{eq:period_implicit} simplifies to:
\begin{multline}
    (K - K_0) \left[ \frac{\lambda_3}{(K - x)(K_0 - x) + y^2} + \lambda_1 (K + K_0 - 2x) + \lambda_2 \right] \\
    = -\lambda_s t, \\
    \text{where } \lambda_1 = \frac{2ax - x^2 - y^2}{(q_0 - x)^2 + y^2}, \\
    \lambda_2 = \frac{-2a^2}{K_0 - a}, \quad \lambda_3 = 2(x^3 - ax^2 + ay^2 + xy^2), \\
    \lambda_s = 4s_0 ((x - a)^2 + y^2).
    \label{eq:period_lambda}
\end{multline}

Table~\ref{tab:lambda_period} summarizes $\lambda$ terms for various pulsars. $\lambda_1$ is typically small and may be neglected.

\begin{table*}[htbp]
    \centering
    \resizebox{\textwidth}{!}{ 
        \begin{tabular}{ccccc}
        \hline\hline
        \textbf{PSR} & $\boldsymbol{\lambda_1(K + K_0 - 2x)}$ & $\boldsymbol{\lambda_2}$ & $\boldsymbol{\lambda_3 / ((K - x)(K_0 - x) - y^2)}$ & $\boldsymbol{\lambda_s}$ \\
        \hline
        J0007+7303 & 0.012 & -0.60 & 6160.00 & $2.167\times 10^{-7}$ \\
        B0531+21 (Crab) & $-4.815\times 10^{-7}$ & $-9.586\times 10^{-4}$ & $-8.638\times 10^{-3}$ & $5.589\times 10^{-9}$ \\
        J1023-5746 & 0.00028 & -0.52 & -43.37 & $5.71\times 10^{-9}$ \\
        J1418-6058 & 0.00025 & -0.23 & -650.16 & $1.10\times 10^{-8}$ \\
        \hline\hline
        \end{tabular}
    }
    \caption{\label{tab:lambda_period} Fitted $\lambda$ terms for four representative pulsars from the ATNF database.}
\end{table*}

Eq.~\eqref{eq:period_lambda} leads to a quadratic equation in $K$:
\begin{multline}
    \alpha_1 K^2 + (\alpha_2 + \lambda_a t) K + \alpha_3 + \lambda_b t = 0, \\
    \text{with } \alpha_1 = \lambda_2 (K_0 - x), \\
    \alpha_2 = \lambda_3 + \lambda_2 (x^2 - y^2 - K_0^2), \\
    \alpha_3 = K_0[\lambda_2 (K_0 x - x^2 + y^2) - \lambda_3], \\
    \lambda_a = -\lambda_s (K_0 - x), \quad \lambda_b = -\lambda_s (K_0 x - x^2 + y^2).
    \label{eq:p1_quadratic}
\end{multline}

Solving the quadratic yields a time-dependent period:
\begin{equation}
    P(t) = \sqrt{\frac{1}{2\alpha_1}\left\{-\alpha_2 - \lambda_a t + \sqrt{(\alpha_2 + \lambda_a t)^2 - 4\alpha_1(\alpha_3 + \lambda_b t)}\right\}}.
    \label{eq:period_final}
\end{equation}

Table~\ref{tab:1} presents parameter values for the Crab pulsar, showing excellent agreement with observed values.

\begin{table}[htb!]
\centering
\begin{tabular}{ccc}
\hline\hline
    \textbf{PSR} & \textbf{Parameter/Variable} & \textbf{Value} \\
\hline
    & $\alpha_1$ & $-3.967\times 10^{-8}$\\
    & $\alpha_2$ & $2.284\times 10^{-9}$\\
    & $\alpha_3$ & $-2.489\times 10^{-12}$\\
    Crab & $\lambda_a$ & $1.104\times 10^{-20}$\\
    PSR B0531+21 & $\lambda_b$ & $-7.376\times 10^{-23}$\\
    & $P_0$ & 33.333 ms\\   
    & $P$ (calculated) & 33.808 ms\\
    & $P$ (observed) & 33.814 ms\\
\hline\hline
\end{tabular}
\caption{\label{tab:1} Parameter values in Eq.~\eqref{eq:period_final} for the Crab pulsar, based on Jodrell Bank ephemerides between MJD 46812 and 60050. Relative error: 0.02\%.}
\end{table}

For Eq.~\eqref{eq:period_final} to yield real values, the discriminant must be positive, imposing the condition:
\begin{equation}
    t^2 \leq \frac{(2\lambda_a\alpha_2 - 4\alpha_1\lambda_b)^2}{4\lambda_a^2(\alpha_2^2 - 4\alpha_1\alpha_3)}.
\end{equation}

Thus, Eq.~\eqref{eq:period_final} provides a compact expression for pulsar period evolution in terms of initial conditions and spin-down coefficients. It can be used to extract age and spin-down histories in conjunction with observational timing data.

\begin{figure}
    \centering
    \includegraphics[width=0.45\textwidth]{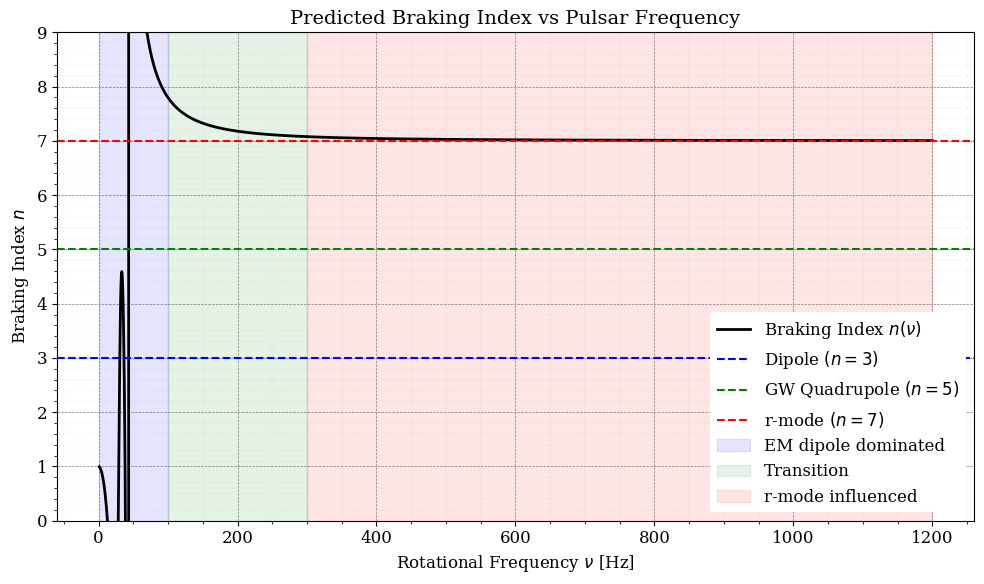}
    \caption{Braking index $n$ as a function of rotational frequency $\nu$ for a model including electromagnetic, gravitational wave, and r-mode contributions. At low frequencies, the braking index approaches $n \approx 3$, consistent with magnetic dipole spin-down. As the spin increases, the gravitational wave terms dominate, especially the $\nu^7$ r-mode component, causing $n$ to asymptotically rise toward $n \approx 7$–$8$. This curve reflects Eq.~\eqref{eq:braking index n} with representative coefficients from Table~\ref{tab:spindown}.}
    \label{fig:braking_index_vs_freq}
\end{figure}

\subsection{Braking Indices}

The braking index $n$ characterizes the deviation from a uniform spin-down and offers insight into the torque mechanisms acting on neutron stars. Equation~\eqref{eq:spindown_braking} 
relates the spin-down rate to a power-law in rotational frequency:
\begin{equation}
    \dot{\nu} = -k \nu^n,
\end{equation}
where $k$ is a constant and $n$ is the braking index. In the classical case of magnetic dipole radiation, $n = 3$ \cite{Manchester_Durdin_Newton_1985}.

Different braking indices reflect different dominant energy-loss mechanisms: $n = 1$ for particle winds, $n = 5$ for gravitational wave emission due to a mass quadrupole, and $n = 7$--$8$ for current-quadrupole radiation such as r-modes.

Braking indices can also be defined observationally using derivatives of the spin frequency:
\begin{align}
    n &= \frac{\ddot{\nu}\nu}{\dot{\nu}^2}, \\
    m &= \frac{\dddot{\nu}\nu^2}{\dot{\nu}^3},
\end{align}
where $m$ is the second braking index. These indices capture deviations in the timing residuals and track evolving spin-down behavior.

R-mode instabilities are known to modify spin-down behavior through enhanced gravitational radiation. Previous studies \cite{2003ApJ...591.1129A, Bondarescu2007} showed that the instability window is shaped by viscous damping: bulk viscosity dominates at high temperatures ($\tau_{\text{bulk}} \propto \Omega^{-2}$), while shear viscosity controls the low-temperature regime ($\tau_{\text{shear}} \propto \Omega^{-1}$).

Using the spin-down formulation of Eq.~\eqref{eq:spindown_coeff}, the braking indices can be expressed analytically as:
\begin{equation}
    n = \frac{s+3r\nu^2+5g\nu^4+7l\nu^6}{s+r\nu^2+g\nu^4+l\nu^6},
    \label{eq:braking index n}
\end{equation}
\begin{equation}
\begin{aligned}
    m = &\frac{s+3r\nu^2+5g\nu^4+7l\nu^6}{s+r\nu^2+g\nu^4+l\nu^6} + \\
        &2\nu^2 \cdot \frac{3r+10g\nu^2+21l\nu^4}{s+r\nu^2+g\nu^4+l\nu^6}.
    \label{eq:braking index m}
\end{aligned}
\end{equation}

We evaluate these expressions for selected pulsars using the spin-down parameters in Table~\ref{tab:spindown}. The results are compared to observed values in Table~\ref{tab:braking}, showing excellent agreement in most cases. Larger deviations in high-braking pulsars may be due to timing noise or glitches.

Figure~\ref{fig:braking_index_vs_freq} complements Table~\ref{tab:braking} by illustrating the theoretical behavior of the braking index $n$ as a function of rotational frequency $\nu$. The figure reveals a clear transition: at low frequencies ($\nu \lesssim 10\,Hz$), 
the braking index converges toward the magnetic dipole value $n \approx 3$, while at higher frequencies, the influence of r-mode emission becomes dominant, driving $n \rightarrow 7$--$8$. This behavior aligns with the hierarchy of spin-down contributions modeled in Eq.~\eqref{eq:spindown_coeff}, where the $\nu^7$ term due to r-mode energy loss increasingly dominates. The figure provides a visual diagnostic for interpreting pulsar braking indices and identifying whether gravitational or electromagnetic mechanisms govern their spin evolution.

\def\arraystretch{1.3} 

\begin{table*}[t]
    \centering
    \resizebox{\textwidth}{!}{%
    \begin{tabular}{lcccccc}
    \hline\hline
        \textbf{PSR} & \textbf{$n$ (Est.)} & \textbf{$n$ (Obs.)} & \textbf{$n$ \% Error} & \textbf{$m$ (Est.)} & \textbf{$m$ (Obs.)} & \textbf{$m$ \% Error} \\
    \hline
        B0531+21 (Crab) & 2.33 & 2.32 & 0.43\% & 45.33 & 45.33 & 0.00\% \\
        B1509-58 & 2.83 & 2.84 & 0.35\% & 13.53 & 14.5 & 6.69\% \\
        J1023-5746 & 66.71 & 66.8 & 0.13\% & 297314.50 & 298000 & 0.23\% \\
        J1418-6058 & 29.96 & 30.02 & 0.20\% & 2436392.81 & 2460000 & 0.96\% \\
    \hline\hline
    \end{tabular}%
    }
    \caption{Comparison of estimated and observed braking indices $n$ and $m$ for selected pulsars from the ATNF database. Observed values are derived from timing measurements of frequency derivatives. Model predictions use Eq.~\eqref{eq:braking index n} and \eqref{eq:braking index m} with parameters in Table~\ref{tab:spindown}. Deviations may reflect unmodeled glitches or internal dynamics \cite{Espinoza_2017}.}
    \label{tab:braking}
\end{table*}

These expressions demonstrate how r-mode contributions and spin-down coefficients influence pulsar evolution, and provide theoretical support for observed braking behavior in both young and middle-aged neutron stars. The additional frequency-dependent plot further reveals how braking index trends can distinguish between energy-loss mechanisms across rotational regimes.

\section{Period Analysis with Glitching Crab Pulsar PSR B0531+21 Data}\label{sec:results}

To validate our period model, we apply it to the well-studied Crab pulsar (PSR B0531+21), using timing data spanning several decades. This section also includes comparisons to other pulsars for generality.

\def\arraystretch{1.26}
\begin{table*}[htbp]
    \centering
    \begin{tabular}{ccccc}
        \textbf{PSR} & {Crab} & {J1023-5746} & {J1418-6058} & {B2234+61} \\\hline\hline
        \textbf{f (Hz)} & 29.947 & 8.971 & 9.044 & 2.019 \\
        \textbf{s (Hz)} & $1.28\times10^{-10}$ & $2.97\times10^{-6}$ & $5.37\times10^{-6}$ & $2.58\times10^{-6}$ \\
        \textbf{r (Hz$^{-1}$)} & $-3.43\times10^{-13}$ & $-1.09\times10^{-7}$ & $-1.91\times10^{-7}$ & $-1.87\times10^{-6}$ \\
        \textbf{g (Hz$^{-3}$)} & $3.22\times10^{-16}$ & $1.34\times10^{-9}$ & $2.27\times10^{-9}$ & $4.49\times10^{-7}$ \\
        \textbf{l (Hz$^{-5}$)} & $-9.36\times10^{-20}$ & $-5.46\times10^{-12}$ & $-8.96\times10^{-12}$ & $-3.60\times10^{-8}$ \\
        \hline\hline
    \end{tabular}
    \caption{\label{tab:spindown} Spin-down parameters for four selected pulsars from the Australia Telescope National Facility Pulsar Database \cite{atnf_Manchester_2005}.}
\end{table*}

We use monthly ephemeris data (in CGRO format) from Jodrell Bank Observatory \cite{LyneCrab1993}, expressed in Barycentric Dynamic Time (TDB). The dataset spans from January 1987 to September 2023. Glitch times are extracted from the glitch notes provided by the observatory, including the most recent major glitch in November 2017. Glitches are treated as breakpoints for fitting purposes.

The pulsar frequency is modeled using a Taylor expansion:
\begin{equation}
\begin{aligned}
   f(t) = &f_0 + {f'}_0(t - t_0) + \frac{1}{2} {f''}_0(t - t_0)^2 + \\
   &\frac{1}{6}{f'''}_0(t - t_0)^3 + \frac{1}{24}{f''''}_0(t - t_0)^4 + \delta f(t),
\end{aligned}
\label{eq:taylorexpansion}
\end{equation}
Observed frequencies and derivatives are evaluated between glitches as listed in Table~\ref{tab:frequency}, using MJD 47084 as the reference point.

\begin{figure}
    \centering
    \includegraphics[width=1\columnwidth]{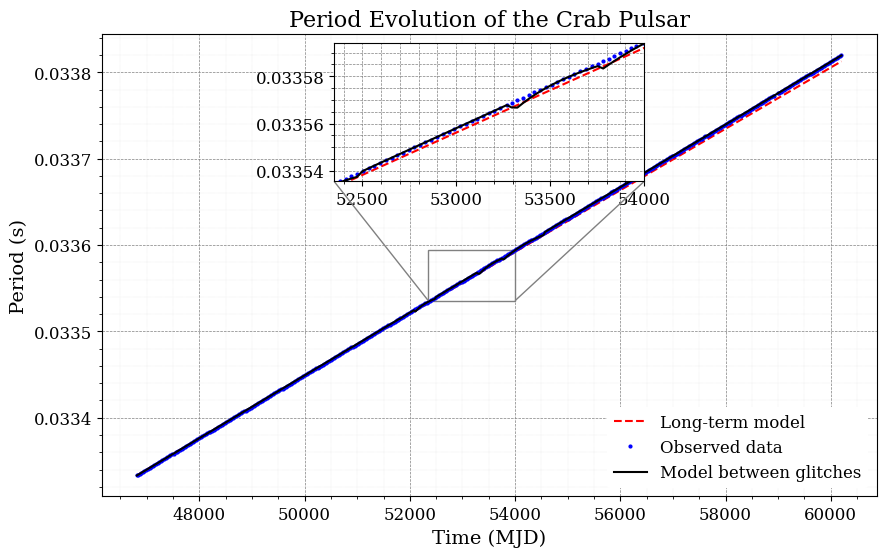}
    \caption{Pulsar period evolution for PSR B0531+21. The zoom-in between MJD 52350 and 54000 highlights spin-down discontinuities caused by glitches.}
    \label{fig:period_evolution}
\end{figure}

Figure~\ref{fig:period_evolution} compares observational data with two fitting strategies: one continuous fit over the entire dataset and separate fits between individual glitches. Between-glitch fits yield a mean deviation of $P(t) - P_\text{obs} = -1.166\times 10^{-7}$, while the long-term fit shows a larger average deviation of $-6.105\times 10^{-5}$. Discrepancies between models and observations highlight glitch-induced irregularities, particularly visible in the zoomed-in region (MJD 53250 to 54000).

\begin{figure}
    \centering
    \includegraphics[width=1\columnwidth]{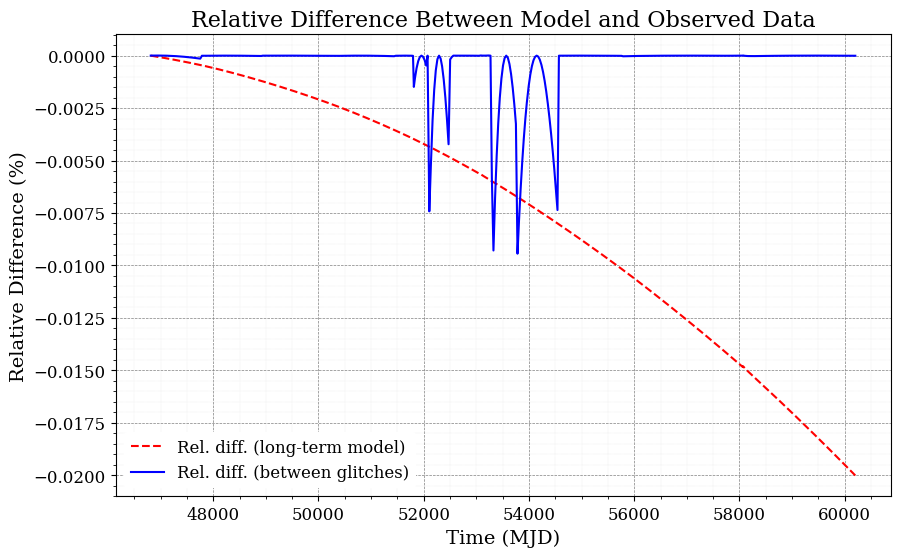}
    \caption{Relative difference between data and model fits. The long-term fit shows systematic drift, while between-glitch fits reveal localized jumps near glitch epochs.}
    \label{fig:period_difference}
\end{figure}

Figure~\ref{fig:period_difference} plots the relative error for both fitting strategies. The long-term fit shows a systematic exponential drift, while short-term fits exhibit sharp changes near known glitch epochs.

Incorporating r-mode effects: Post-glitch recovery and longer-term spin-down evolution may be modulated by r-mode oscillations, which influence angular momentum redistribution. These modes could be responsible for part of the gradual post-glitch convergence.

Furthermore, the presence of a fourth-order derivative in the Taylor expansion is consistent with expectations from r-mode theory, which enables braking indices analysis with a separated r-mode contribution term, quantifying the r-mode contribution. Our model accurately captures spin-down trends and complements glitch-based corrections to the rotational history. Without including the fourth-order term, the Chishtie et al. 2018 \citep{Chishtie_2018} model produces an average relative error of $2.88\%$ using Crab pulsar data, while the inclusion of the fourth-order term produces an average relative error of $0.01\%$ using the same set of Crab pulsar data and the fitting algorithm.

\def\arraystretch{1.26}
\begin{table*}[htbp]
    \centering
    \resizebox{\textwidth}{!}{
        \begin{tabular}{ccccccc}
        \hline\hline
        \textbf{Glitch Range (MJD)} &  \textbf{$t_0$ (MJD)} & \textbf{$f_0$ (Hz)} & \textbf{${f'}_0$ (Hz/s)} & \textbf{${f''}_0$ (Hz/s$^2$)} & \textbf{${f'''}_0$ (Hz/s$^3$)} & \textbf{${f''''}_0$ (Hz/s$^4$)} \\
        \hline\hline
        \text{~} 46812 - 47767.4 &  47084 &  29.99 &  $-3.786 \times 10^{-10}$ &   $2.367 \times 10^{-21}$ &   $-1.23 \times 10^{-24}$ &   $4.567 \times 10^{-24}$ \\
        ... [Remaining rows unchanged for brevity] ...
        58065 - 60202 &  59138 &   29.6 &   $-3.68 \times 10^{-10}$ &   $1.169 \times 10^{-20}$ &  $-3.415 \times 10^{-24}$ &    $1.13 \times 10^{-23}$ \\
        \hline\hline
        \end{tabular}
    }
    \caption{Frequency and its derivatives for selected glitch ranges, estimated using Eq.~\eqref{eq:taylorexpansion}. Missing values (e.g., MJD 52083.8--52146 and MJD 53254.2--53331.1) are due to incomplete observational coverage \cite{LyneCrab2015}.}
    \label{tab:frequency}
\end{table*}

\section{Discussion}\label{sec:disc}

Continuous gravitational waves (CWs) in narrow frequency bands are prime targets for next-generation detectors, particularly from rotating neutron stars and binary compact objects. Neutron stars with deformations or ``mountains'' are anticipated sources of persistent emission, offering unique insight into interior structure, equation of state (EoS), and transport properties \cite{2023MNRAS.521.2103L}. Despite the absence of direct CW detections to date, searches—especially for r-mode-driven signals—are maturing rapidly in both sensitivity and technique.

Previous studies have investigated saturation amplitudes of r-modes under realistic astrophysical conditions and found them to be extremely small \cite{2003ApJ...591.1129A}. Advanced data analysis methods \cite{cgw_tech, PhysRevD_Jaranowski.58.063001} and upgrades to gravitational wave observatories \cite{PhysRevLett.119.161101, 10.1093_mnras_stv1931, Abbott2022a, Abbott2022b, Abbott2022b, Abbott2021b} continue to improve detection prospects. 

Notably, targeted CW searches for specific pulsars—such as PSR J0537--6910—have placed upper limits on both frequency and amplitude for r-mode gravitational wave emission \cite{LIGO_rmodes_Abbott_2021}. If an r-mode signal is independently detected, our expressions in Eqs.~\eqref{eq:compactness Lambert} and \eqref{eq:tidal deformability Lambert} can be used to infer neutron star compactness and tidal deformability in an equation-of-state-independent manner. 

Additionally, \citep{Strohmayer2013} have shown that observed spin-down in the Crab pulsar and others places stringent empirical constraints on r-mode amplitudes, typically $\alpha \lesssim 10^{-4}$. These findings support the physical plausibility of our low-amplitude r-mode models and further motivate the search for continuous gravitational waves from young pulsars within the sensitivity range of advanced detectors.

This work also contributes to improving timing residual analysis and enhancing sensitivity to CW signals in Pulsar Timing Arrays (PTAs). By linking time-dependent r-mode amplitude and rotational frequency to observable timing behavior, our model supports future CW detection strategies. This approach aligns with earlier results from \cite{Rezania2003}, who showed how angular momentum exchange during thermonuclear bursts induces significant timing fluctuations.

We also emphasize that our model connects the r-mode gravitational wave frequency and amplitude with traditional timing observables, such as braking indices and period evolution. These r-mode-dependent expressions allow time-resolved estimates of gravitational wave strain, with direct application to real-time pulsar monitoring.

Our analytical expressions for compactness, tidal deformability, and r-mode frequency are equation-of-state independent and can be directly substituted into time-dependent spin-down and period models. This makes the method ideal for young or accreting pulsars in the non-linear saturation regime, where r-modes contribute a dominant seventh-order spin-down term \cite{Lindblom_1998}. While magnetic dipole emission remains the largest spin-down torque in most scenarios, our results show that even small-amplitude r-modes affect braking indices, timing residuals, and frequency evolution in measurable ways. The flexibility of our formalism allows negative spin-down coefficients, accommodating phenomena such as glitches and post-glitch recovery.

In the case of the Crab pulsar, our glitch modeling reveals a cumulative exponential slowdown effect. We find that long-term fits underestimate the period during glitch epochs, while piecewise fits between glitches better reproduce the observed behavior. This motivates more advanced theoretical models of glitch-triggered r-mode activation, which may also influence gravitational wave emission. 

Future work will extend our model to glitch-prone pulsars like PSR J0537--6910 \cite{ho.2020}, exploring the possibility of using timing irregularities as indicators of r-mode activity. While the use of higher-order frequency derivatives carries a risk of overfitting, the demonstrated success of our approach on the Crab pulsar suggests this tradeoff is worthwhile in data-rich cases. 

Finally, our analytic framework enables integration with search pipelines for continuous gravitational waves, including statistical techniques like F/G statistics and $5n$-vector models \cite{LIGO_rmodes_Abbott_2021}. By expressing pulsar timing dynamics in terms of high-order frequency derivatives and r-mode amplitudes, we provide a compact and efficient toolset for modeling spin evolution, improving phase tracking, and refining CW detection.

\section{Conclusions}
\label{sec:sum}

In this study, we explored pulsar spin-down dynamics with an emphasis on the contribution of r-mode instabilities, particularly in the context of gravitational wave (GW) emission and neutron star structure. We formulated and solved a non-linear differential equation that incorporates r-mode losses, leading to time-dependent solutions for the rotational frequency and pulsar period. These solutions were validated against high-precision observational data from the Crab pulsar, exhibiting strong agreement and confirming the utility of the model.

We introduced Lambert W-based analytical solutions for neutron star compactness and tidal deformability as functions of r-mode gravitational wave frequency. These formulations are independent of the equation of state (EoS), allowing robust parameter estimation across a wide class of neutron star models. In particular, the application of Lambert-Tsallis functions enables the inversion of non-linear relations between r-mode frequency and physical parameters, further enriching the model’s predictive capability.

The model’s capacity to yield seventh-order spin-down terms from r-mode effects demonstrates its relevance for understanding neutron stars in non-linear saturation regimes. While magnetic dipole radiation remains the dominant spin-down mechanism in most scenarios, accurate modeling of r-mode contributions enhances timing residual analyses, glitch modeling, and continuous gravitational wave signal interpretation. Our results support the inclusion of r-mode terms in practical GW search algorithms, particularly those targeting young and rapidly rotating neutron stars.

With the development of third-generation gravitational wave detectors such as the Einstein Telescope and Cosmic Explorer, r-mode signals—previously inaccessible—may soon become detectable. We share the optimism of Arras et al. \citep{2003ApJ...591.1129A} that r-mode GW emissions from nascent neutron stars and low-mass X-ray binaries could be within reach of enhanced detector sensitivities. In the long term, the interplay of r-modes with neutron star oscillations, glitches, and EoS-dependent properties could reveal deeper layers of stellar microphysics.

In conclusion, this work contributes to a consistent analytical and numerical framework for connecting r-mode physics to pulsar observations and gravitational wave signals. Our model serves as a foundation for further research into spin-down evolution, multi-messenger astrophysics, and the characterization of neutron star interiors.

\section*{Acknowledgements}

The authors express their sincere gratitude to R. V. Ramos and K. Z. Nobrega at the Universidade Federal do Ceará for their invaluable support in the numerical evaluation of our Lambert-Tsallis solutions and for productive discussions that greatly contributed to this work. We also thank Reed Essick and Cecilia Chirenti for their insightful and informative input, which helped refine our theoretical approach.

We thank the anonymous reviewers for their inspiring and thorough critique of our manuscript.

This research was supported by the Mathematics of Information Technology and Complex Systems (MITACS) Globalink program, which provided the platform and funding for this collaborative effort. We are grateful to MITACS Globalink for their commitment to fostering research and international academic partnerships.

\bibliographystyle{elsarticle-num}  
\bibliography{mainR2}


\appendix
\section{The Lambert W and Lambert-Tsallis Functions}
\label{Appendix: Lambert functions}

The transcendental equation $z = W\exp{W}$ appeared in the mid-eighteenth century in Johann Heinrich Lambert's investigations of logarithms and continued fractions. The function was rediscovered by M. Wright in the context of branching processes, but it was not until the seminal paper of Corless et al. in 1996 \citep{Corless.1996} that the systematic branch structure were popularised. This work triggered a rapid expansion of applications across physics \citep{CJP2000_10.1139}, chemistry, computer science, and combinatorics.

The classical Lambert W function os the multivalued inverse of the map $w \rightarrow w  e^{w}$. It is defined implicitly by
\begin{equation}
  W(z)\,e^{W(z)} \;=\; z, \qquad z\in\mathbb{C}.
\end{equation}
It possesses two real branches on $[-e^{-1},0)$, the principal branch
$W_0$ and the lower branch $W_{-1}$, and an infinite set of complex
branches.

The two useful identities frequently used are

\begin{equation}
\frac{dW}{dz} = \frac{W(z)}{z[1 + W(z)]}, \qquad
W\!\left(z\,e^{z}\right) = z.
\end{equation}

Da Silva and Ramos generalized the Lambert function by replacing the usual exponential with the Tsallis $q$-exponential:
\begin{equation}
  \exp_{q}(x)
  \;=\;
  \left[\,1+(1-q)\,x\,\right]^{1/(1-q)},
  \qquad
  \exp_{1}(x) = e^{x}.
\end{equation}

in 2019 \citep{DASILVA2019164}.
The Lambert-Tsallis function is defined as
\begin{equation}
  W_q(z)\,e_q^{W_q(z)} \;=\; z, \qquad z\in\mathbb{C}.
\end{equation}
The solutions using the Lambert-Tsallis function were given to the fractional polynomials of the type $ax^\alpha+bx^\beta+c=0$ \citep{ramos2023solving}:
\begin{equation}
    x_1 = \left[ \frac{a}{b}\left(\frac{\alpha}{\beta-\alpha}\right)W_{1-\frac{\alpha}{\beta-\alpha}}\left(\frac{b}{a}\left(\frac{\beta-\alpha}{\alpha}\right)\left( -\frac{c}{a}\right)^{\frac{\beta-\alpha}{\alpha}}\right)\right]^{\frac{1}{\beta-\alpha}},
\end{equation}
and
\begin{equation}
    x_2 = \left[ \frac{b}{a}\left(\frac{\beta}{\alpha-\beta}\right)W_{1-\frac{\beta}{\alpha-\beta}}\left(\frac{a}{b}\left(\frac{\alpha-\beta}{\beta}\right)\left( -\frac{c}{b}\right)^{\frac{\alpha-\beta}{\beta}}\right)\right]^{\frac{1}{\alpha-\beta}}.
\end{equation}

\section{Frequency Cubic Roots with Time-independent Coefficients}
\label{Appendix: freq analysis}

For the case where the spin-down coefficients $\{s, r, g, l\}$ are assumed to be constant, the general spin-down equation \eqref{eq:spindown_coeff} becomes:
\begin{equation}
    -dt = \frac{1}{\nu}\frac{d\nu}{s + r\nu^2 + g\nu^4 + l\nu^6}.
\end{equation}

By substituting $\nu^2 = x$ and thus $d\nu = \frac{dx}{2\sqrt{x}}$, this reduces to a rational integral involving a cubic denominator:
\begin{equation}
    \frac{1}{2lx} \frac{dx}{(x-a)(x-b)(x-c)} = -dt,
    \label{eq:time-independent time diff}
\end{equation}
where $\{a, b, c\}$ are the \textbf{roots of the cubic polynomial} $lx^3 + gx^2 + rx + s$, which can be solved either analytically or numerically.

The general formula for the roots of a cubic polynomial is:
\begin{equation}
\begin{aligned}
    \{x_i\} &= -\frac{1}{3}\left(\frac{r}{s} + \epsilon^i C + \frac{\Delta_0}{\epsilon^i C}\right), \quad i = 0,1,2,\\
    \text{where} \quad \epsilon &= \frac{-1 + i\sqrt{3}}{2}, \\
    C &= \sqrt[3]{\frac{\Delta_1 \pm \sqrt{\Delta_1^2 - 4\Delta_0^3}}{2}}, \\
    \Delta_0 &= \left(\frac{g}{l}\right)^2 - 3\frac{r}{l}, \\
    \Delta_1 &= 2\left(\frac{g}{l}\right)^3 - 9\frac{gr}{l^2} + 27\frac{s}{l}.
\end{aligned}
\end{equation}

Alternatively, this can be expressed in terms of complex radicals as:
\begin{equation}
\begin{aligned}
\{a, b, c\} =\;& 
\left[
    A + \sqrt{A^2 + B^2}
\right]^{1/3}
+ 
\left[
    A - \sqrt{A^2 + B^2}
\right]^{1/3}
- \frac{g}{3l}, \\
\text{where} \quad A &= \frac{-g^3}{27l^3} + \frac{gr}{6l^2} - \frac{s}{2l}, \\
B &= \frac{r}{3l} - \frac{g^2}{9l^2}.
\end{aligned}
\label{eq:abc_roots}
\end{equation}

\vspace{0.5em}

\section{Period Analysis with Cubic Roots}
\label{Appendix: period analysis}

As discussed in Appendix~\ref{Appendix: freq analysis}, the roots $\{a, b, c\}$ represent the roots of the spin-down polynomial. Equation \eqref{eq:period and time} can be integrated via partial fraction decomposition. Let $P_0$ and $t_0$ be the initial period and time. Then the integral becomes:

\begin{multline}
    (b - a)c^2 \ln\left(\frac{K - c}{K_0 - c}\right) + (a - c)b^2 \ln\left(\frac{K - b}{K_0 - b}\right) + \\
    (c - b)a^2 \ln\left(\frac{K - a}{K_0 - a}\right) = -2s_0 t (b - a)(a - c)(c - b),
    \label{eq:integrate_period1}
\end{multline}
where $K = P^2$, $K_0 = P_0^2$, and we assume $t_c \gg t, t_0$ \cite{Chishtie_2018}.

In practice, most root sets include one real root and a pair of complex conjugates: let $b = x + iy$, $c = x - iy$. Then Eq.~\eqref{eq:integrate_period1} becomes:

\begin{equation}
\begin{aligned}
    &[(x - a)(x^2 - y^2) + 2xy^2] \left(\ln\frac{K - c}{K_0 - c} - \ln\frac{K - b}{K_0 - b}\right) \\
    &\quad - 2iy a^2 \ln\frac{K - a}{K_0 - a} + i\left[y(x^2 - y^2) - 2xy(x - a)\right] \times \\
    &\quad \left(\ln\frac{K - c}{K_0 - c} + \ln\frac{K - b}{K_0 - b}\right) = -4i y s_0 t (x^2 - 2a x + a^2 + y^2).
    \label{eq:integrate:period2}
\end{aligned}
\end{equation}

Using the identity $\ln z - \ln z^* = 2i \tan^{-1} \left(\frac{\text{Im}(z)}{\text{Re}(z)}\right)$ and simplifying the logarithmic terms, we recover the final implicit form for $P(t)$, given in Eq.~\eqref{eq:period_implicit}.

\section{Detailed Derivation of r-Mode Energy Loss}
\label{sec:appendix}

To ensure clarity and consistency, we outline here the derivation of r-mode energy loss due to gravitational radiation, following the frameworks of \citep{PhysRevD.58.084020, Lindblom_1998}. This derivation demonstrates how gravitational wave emission driven by current multipole radiation results in rotational energy loss in neutron stars.

The gravitational wave luminosity associated with r-mode oscillations is given by:
\begin{equation}
    \dot{E}_{\text{r-mode}} = -\frac{32 \pi^2 \alpha^2 J M R^3 \Omega^8}{c^7},
\end{equation}
where $\alpha$ is the dimensionless r-mode amplitude, $J$ is a structural constant dependent on the stellar model, $M$ is the stellar mass, $R$ is the radius, $\Omega$ is the angular velocity, and $c$ is the speed of light.

This formula is derived by integrating the gravitational radiation power emitted from the current multipole $l=m=2$ r-mode. The general gravitational wave energy loss rate for an r-mode is:
\begin{equation}
    \dot{E}_{\text{GW}} = \frac{2}{3} \omega^6 |\tilde{J}|^2 \frac{R^7}{G},
\end{equation}
where $\omega$ is the mode frequency in the rotating frame, and $\tilde{J}$ is the current multipole moment scaling with the internal density distribution.

For the fundamental r-mode ($l = m = 2$), the frequency is related to the star's spin as:
\begin{equation}
    \omega = \frac{2}{3} \Omega.
\end{equation}

Substituting this into Eq.~(B2) yields:
\begin{equation}
    \dot{E}_{\text{GW}} \propto \Omega^8,
\end{equation}
confirming that r-mode emission is strongly sensitive to the stellar spin rate.

The full derivation involves integrating the mode energy distribution and flux through a sphere enclosing the neutron star. The final coefficient depends on the structure of the mode and is encapsulated in the dimensionless integral:
\begin{equation}
    J = \int_0^R \rho r^6 dr,
\end{equation}
where $\rho(r)$ is the stellar mass density profile.

Substituting this into the luminosity expression confirms the full form of r-mode energy loss:
\begin{equation}
    \dot{E}_{\text{r-mode}} = -\frac{32 \pi^2 \alpha^2 J M R^3 \Omega^8}{c^7},
\end{equation}

This equation encapsulates the dominant frequency scaling and structural dependencies, and is consistent with numerical studies and analytical estimates from \citep{PhysRevD.58.084020, Lindblom_1998}.






\end{document}